# 3D printing in the context of Science, Technology, Engineering, and Mathematics education at the college/university level

Peter Moeck[1*], Paul DeStefano[1], Werner Kaminsky[2], and Trevor Snyder[3,4]

[1] Nano-Crystallography Group, Department of Physics, Portland State University, Portland, OR 97207-0751, USA,
* pmoeck@pdx.edu
[2] Department of Chemistry, University of Washington, Seattle, WA 98195, USA
[3] 3D Systems Corporation, 26600 SW Parkway, Wilsonville, OR 97070, USA
[4] Department of Mechanical and Materials Engineering, Portland State University, Portland, OR 97207-0751, USA

An overview concerning 3D printing (a.k.a. additive manufacturing) within the context of Science, Technology, Engineering, and Mathematics (STEM) education at the college/university level is provided. The vast majority of quoted papers report self-made models for which faculty members and their students have created the necessary 3D print files themselves by various routes. The prediction by the Gartner consulting company that it will take more than ten years from July 2014 onwards for 'Classroom 3D Printing' to reach its 'Plateau of Productivity' in one of their hallmark 'Visibility versus Time (Hype Cycle)' graphs is critically assessed. The bibliography of this book chapter sums up the state-of-the art in 3D printing for STEM (including nano-science and nano-engineering) education at the college level approximately four years after Gartner's prediction. Current methodologies and best practices of college-level 'Classroom 3D printing' are described in the main section of this review. Detailed information is given mainly for those papers in which the authors of this book chapter are authors and co-authors. A straightforward route from crystallographic information framework files (CIFs) at a very large open-access database to 3D print files for atomic-level crystal and molecule structure models is described in some detail. Only the exported 3D print files are downloaded from the website of our 3D converter tool as no local installation of any supporting program or applet is necessary. Because the development of methodologies and best practices are typical activities of the penultimate stage of a 'Hype Cycle', we conclude that (*i*) Gartner's prediction underestimates the creativity, resourcefulness, and commitment of college educators to their students and that (*ii*) 'Classroom 3D Printing' will be a widespread reality significantly earlier than the middle of the next decade (at least in the USA as more than one half of the relevant/quoted papers originated there). An appendix provides a brief technical review of contemporary 3D printing techniques.

## 1. Introduction

This book chapter reviews 3D printing (a.k.a. additive manufacturing) in the context of Science, Technology, and Mathematics (STEM) education at the college/university level and consists of five sections. The emphasis of this review is on current methodologies and best practices of college-level 'Classroom 3D printing' as defined below by the Gartner consulting company, i.e. the creation of both 3D print files and printed-out models, rather than the pedagogical benefits of using such models in classrooms.

A rather comprehensive bibliography on this topic that goes back to the year 2005 follows this brief introduction (as second section of the book chapter). The size of this bibliography justifies the writing of this review. In the third section of this review, we present a critical assessment of a 'Visibility versus Time (Hype Cycle)' graph from the Gartner consulting company. While that particular graph pertains to 3D printing and dates back to July of 2014, there is also a second graph with an analogous outlay (from Wikipedia) that provides verbal illustrations on how Gartner's hype cycle graphs are to be read. The third section of this review provides also a direct quote by the Gartner consulting company on their definition of 'Classroom 3D printing'.

The fourth and main section of this book chapter describes current 'methodologies and best practices' in 'Classroom 3D printing' for STEM education at the college level. Open access sources for crystallographic information framework files (CIFs) are discussed at the beginning of that section. While not providing many details of this very versatile file format of the International Union of Crystallography (IUCr), we will briefly comment throughout that section on crystallographic concepts that are of importance to the creation of specific 3D print files. Three Microsoft Windows[TM] executable programs for the creation of 3D print files of small molecules and proteins from experimentally obtained CIFs, crystal morphologies from CIFs (where crystal morphology information has been included), and representations of typically anisotropic tensor properties of crystals are reviewed. Other stand-alone computer programs that use CIFs as input and allow for the export of 3D print files are also mentioned.

Open access resources for the creation of 3D print files for bioscientific and biomedical applications as well as representations for anisotropic physical properties of crystals (in the form of tensor surface representations of longitudinal effects) are mentioned as well. A straightforward route from CIFs at the Open-Access Crystallography project (Moeck 2004), which is run by the first author's Nano-Crystallography Group at Portland State University (PSU) to 3D print files for crystal structures and molecules (Moeck 2018) is described here in some detail. Other websites that are backed up by large databases and allow for the creation of 3D print files for STEM education are also discussed in the main section of this book chapter. This is followed by the final summary and conclusions section. An appendix provides a brief technical review of contemporary 3D printing techniques.

## 2. Literature Overview

Over more than a decade, hardware cost and performance improvements as well as complementing software developments have made 3D printing practical for usage by college educators who lecture and run teaching laboratories in the STEM fields. Educational 3D printed models of various kinds have so far been utilized for the visualization of a wide range of concepts in *mathematics* (Hart 2005, Segerman 2012), *computer science* (Papazafiropulos 2016), *crystallography* (Moeck 2014a, Chen 2014, Kaminsky 2014, Kitson 2014, Moeck 2014b, Moeck 2014c, Snyder 2014, Teplukhin 2015, Gražulis 2015, Stone-Sundberg 2015, Kaminsky 2015, Moeck 2017a), *mineralogy* (Moeck 2014a, Kaminsky 2014, Stone-Sundberg 2015, Moeck 2016a, Moeck 2018), *geosciences* (Horowitz 2014), *crystal and condensed matter physics* (Stone-Sundberg 2015, Kaminsky 2015, Moeck 2016a, Casas 2018, Moeck 2018), *structural biology* (Gillet 2005, Roberts 2005, Herman 2006, Bain 2006, Olson 2006, Jittivadhna 2010, Wedler 2012, Moeck 2014a, Violante 2014, Meyer 2015, Stone-Sundberg 2015, Gardner 2016, Davenport 2017, Kat Cooper 2017, Da Veiga Beltrame 2017, Van Wieren 2018, Jones 2018, Paukstelis 2018, Tavousi 2018), *chemistry* (Flint 2011, Wedler 2012, Halford 2014, Moeck 2014b, Kaminsky 2014, Scalfani 2014, Lolur 2014, Blauch 2014, Violante 2014, Teplukhin 2015, Striplin 2015, Casas 2015, Rossi 2015, Rodenbough 2015, Kaliakin 2015, Robertson 2015, Stone-Sundberg 2015, Kaminsky 2015, Chen 2015, Smith 2016, Griffith 2016, Smiar 2016, Scalfani 2016a, Moeck 2016a, Wood 2017, Higman 2017, Penny 2017, Carroll 2017, Piunno 2017, Jones 2018, Brown 2018, Paukstelis 2018, Moeck 2018), *materials science and engineering* (Moeck 2014a, Moeck 2014b, Moeck 2014c, Rodenbough 2015, Gražulis 2015, Van Wieren 2018), *nano-science and engineering* (Moeck 2014c, Scalfani 2015, Van Wieren 2018, Jones 2018, Tavousi 2018), and *mechanical engineering* (Snyder 2014, Chen 2015, Gatto 2015). Many of the above quoted papers offer copies of the created 3D print files in their electronic on-line supporting material. More than half of these papers originated at colleges and universities that are located within the United States of America (USA).

As a genuinely interdisciplinary concept, 3D printed visualizations of 3D point symmetries are provided by Flint 2011, Scalfani 2014, Casas 2015, Stone-Sundberg 2015, Moeck 2016a, and Wood 2017. Except for Moeck 2016a, no attempt has been made in these papers to discuss 3D point symmetries in connection with the physical properties of crystalline materials. In addition to giving an overview of 3D printing for STEM education and displaying more than thirty 3D printed models from six different research groups, the paper by Stone-Sundberg 2015 presents a brief history of crystallographic information representations and provides technical commentaries on several 3D printing technologies.

The paper by Moeck 2014c demonstrates 3D printed models that find good uses in ongoing nano-materials science and engineering, materials physics, and introductory nano-science and nano-technology classes at a research university at both the graduate and undergraduate levels. A null-hypothesis test demonstrated (in a qualitative manner) the benefits to the undergraduate students' comprehension and retention of the discussed materials when 3D printed models were used in an introductory nano-science and nano-technology classroom (Moeck 2014c).

Approximately one half of the above-mentioned papers have been published in the *Journal of Chemical Education* of the American Chemical Society. Crystallography journals also helped to bring the current state of the art in 3D printing to the attention of college educators. Since the vast majority of the above-mentioned papers are not specifically about educating visually impaired or blind students (whereby Wedler 2012 and Papazafiropulos 2016 are the only exceptions), it is fair to state that 3D printing has entered mainstream college education in the STEM fields.

The usage of a 3D printing pen for the creation of pseudo-3D models and their utilization to teach the valence shell electron pair repulsion concept to undergraduate chemistry students was reported by Dean 2016. That particular report differs from the common theme of the above-mentioned references in this section of this book chapter significantly so that it is mentioned here separately. In each of the cases of the former references, 3D models existed in a computer before the corresponding 3D print files were sent to 3D printers. In Dean 2016, on the other hand, the 3D models existed originally only in the mind of the instructor and emerged later on in the minds of the students when they assembled various pseudo-2D bits and pieces into various pseudo-3D models with 3D printing pens. Utilizing ordinary cardboard cutouts and some glue would obviously have been a cost and time effective alternative to the teaching and learning activities described in Dean 2016.

A large body of literature on 3D printing in support of college education within the arts, architecture, humanities, human/animal/plant anatomy and social sciences exist also. We are to a large extent ignorant about that literature, but suspect that most readers of this book chapter do not expect to be informed about interesting developments in non-STEM education.

The technical review by Gross 2014 sums up the state-of-the-art of 3D printing and offers an appraisal of its future directions (including applications within college education) when 3D printers become even more accessible. Snyder 2014 provides an overview over 3D Systems Corporation's technologies in connection with new applications of 3D printing for STEM education as well as in Science, Engineering, and Manufacturing. Nadel 2017 reports on different ways to provide 3D printing capabilities for students and faculty members alike at two US American research universities and a School of Design in midtown Manhattan, New York (New York, USA). A preliminary assessment of some health and safety concerns during 3D printing, i.e. the emissions of unwanted particles into the air in the library of a university, has been provided by Bharti 2017.

A summary of important results of a workshop that was sponsored by the National Science Foundation (NSF of the USA) and its recommendations concerning 3D printing within the context of STEAM, i.e. STEM plus the arts, education is provided by Simpson 2017. The state-of-the-art and future potential of 3D printing in general are reviewed by Huang 2015 at an advanced technical level in a paper that resulted from another NSF sponsored workshop. A review of the "where and how of 3D printing" in the wider educational system was published recently by Ford 2018. That review contains references to 328 academic papers, but does not provide much detailed practical information that could guide interested STEM educators in their creation of specific 3D printed models.

### 3. Predictions about the future of 3D printing for STEM/STEAM education at the college level

Figure 1 shows predictions about the future development of 3D printing by the Gartner consulting company in one of their hallmark 'Expectations versus Time (Hype Cycle)' plots from the months of July of the year 2014.

Gartner's definition of 'Classroom 3D Printing' is: "*the practice of using a device to create physical objects from digital models for the purpose of creating teaching aids or demonstrating a concept in 3D. The printer converts a 3D model — for example, created through a computer-aided design or modeling software package — into a solid, detailed and potentially functional model*" (Van der Meulen 2016). This definition covers effectively the topics of all of the above quoted papers, except for the ones by Gross 2014, Huang 2015, Dean 2016, Nadel 2017, Bharti 2017, Simpson 2017, and Ford 2018. It was also made clear by Gartner that the teaching aids and demonstrations of concepts in 3D refer in their definition mainly to college/university classrooms (Van der Meulen 2016).

Figure 2 provides another hype cycle graph (Wikipedia 2018), but instead of focusing on a specific technology, it provides verbal illustrations on how Gartner's hallmark graphs are to be read. Because there is no scientific foundation that underlies Gartner's hype cycle graphs (Wikipedia 2018), these kinds of graphs are simply supposed to be visual aids for the appreciation of 'Amara's Law'.

Roy Amara's approximately five decades old observation states that people generally tend to overestimate the effects of a new technology in the short run and underestimate its effects in the long run (Ridley 2017, Wikipedia 2018). This is thought to create 'Inflated Expectations' in the short run so that a subsequent sliding into a 'Trough of Disillusionment' is unavoidable. Really useful innovations/technologies will in the long run reemerge from this trough and eventually reach their 'Plateau of Productivity'. Some other innovations/technologies may, however, become obsolete before they ever reach this plateau. The projected time spans, i.e. 'less than 2 years', '2 to 5 years', '5 to 10 years', and 'more than 10 years' for a particular innovation/technology to reach its 'Plateau of Productivity' are represented on a generic hype cycle curve by color coded circular and triangular markers, see Fig. 1.

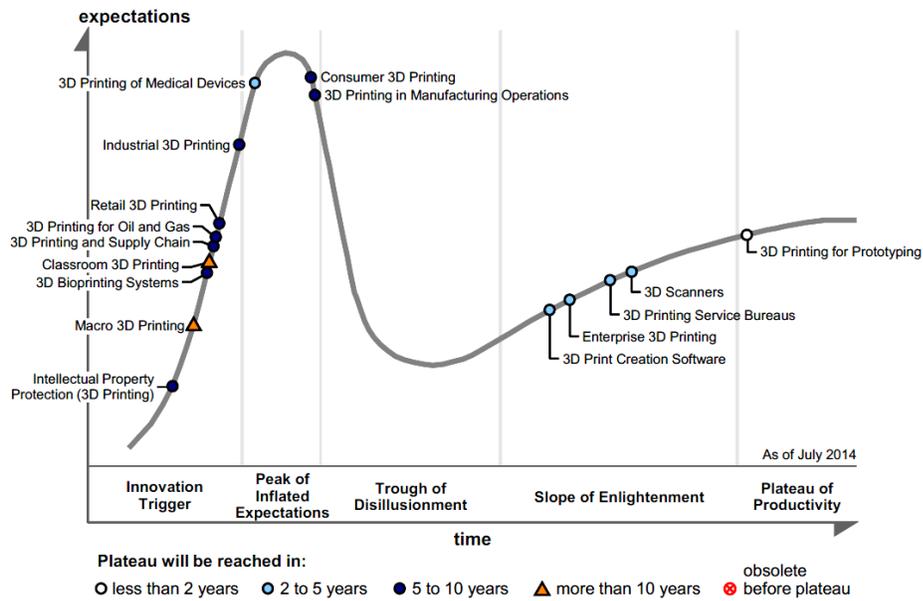

**Figure 1:** Gartner's Hype Cycle for 3D Printing, July 2014. (With Permission by Gartner.)

According to Fig. 1, it will take 'Classroom 3D Printing' more than 10 years to reach its 'Plateau of Productivity'. That stage is the last one of the five stages of any hype cycle and according to Fig. 2 characterized by the fact that '20% to 30% of the potential audience has adopted the innovation'. In July of 2014, 'Classroom 3D Printing' was according to Gartner in the first of these five stages, i.e. the 'Innovation/Technology Trigger' phase, see Fig. 1.

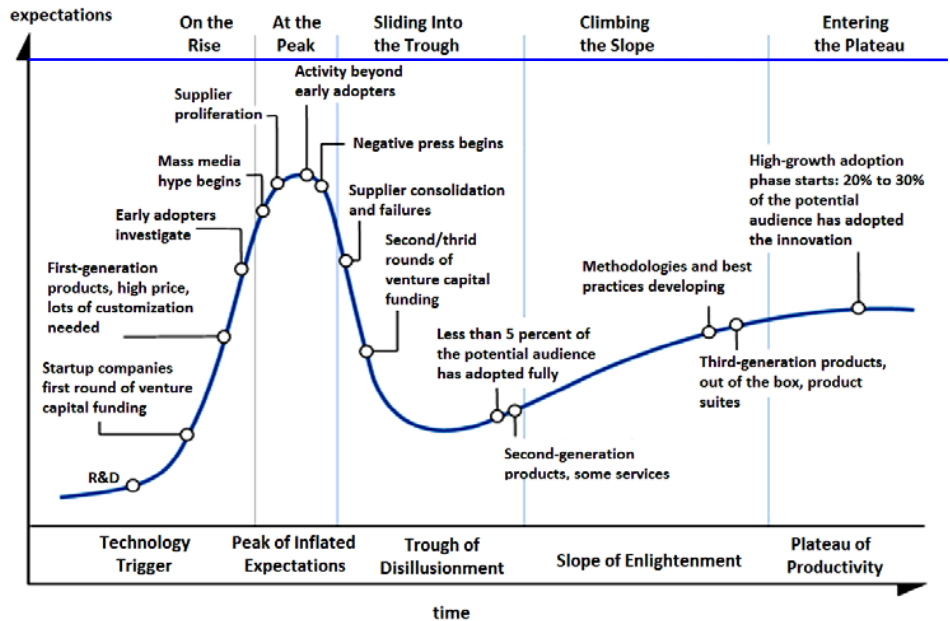

**Figure 2:** How to read a hype cycle graph according to Wikipedia 2018. (Slightly modified version of the graph at https://commons.wikimedia.org/w/index.php?curid=27546041, where it was provided with a CC BY-SA 3.0 license.)

With respect to many of Gartner's other predictions around 3D printing, see Fig. 1 from July 2014, it is fair to state that there is ample anecdotal evidence that '3D Printing of Medical Devices', '3D Print Creation Software', 'Enterprise 3D Printing', '3D Printing Service Bureaus', and '3D Scanners' have by now all progressed to the final stage of the hype cycle. All of these characteristics are in Fig. 1 correspondingly marked with light blue circles for prospectively reaching their 'Plateau of Productivity' in 2 to 5 years.

As far as the above quoted papers that pertain to college-level 'Classroom 3D printing' are concerned, the largest fractions of them appeared in the years 2014 and 2015, not many before that, and smaller quantities since then in the years 2016 and 2017 as well as in the first half of 2018. From this, one might conclude that the 'Peak of Inflated Expectations', i.e. the second stage of the hype cycle, had been reached sometimes around the years 2014 and 2015 and that the technology might currently be in the 'Trough of Disillusionment', i.e. the third of the five cycle stages. According to Fig. 2, this trough is characterized by the fact that 'less than 5 percent of the potential audience has adopted' the new technology.

As there is no scientific foundation for hype cycle graphs, one may as well conclude that 'Classroom 3D printing' is currently on its technological 'Slope of Enlightenment', i.e. the penultimate stage of the cycle, as 'Methodologies and best practices' are 'developing', see Fig. 2. Such methodologies and best practices are the subject of the following (and main) section of this book chapter. Due to the quite mature state of the affairs that is described in that particular section, we conclude that Gartner's prediction underestimates the commitment of college STEM educators to their students as well as their creativity and resourcefulness. 'Classroom 3D printing' at the college level may, therefore, reach its 'Plateau of Productivity' significantly earlier than the middle of the next decade (at least in the USA). The academic freedom that comes with tenured faculty positions at colleges/universities (in the USA) will surely trump college bureaucracies in making this happening.

The review by Huang 2015 features a hype cycle graph pertaining to 3D printing/additive manufacturing in general. According to that graph (Huang 2015), which is in line with predictions by Gartner, the additive manufacturing field, as a whole, was at its 'Peak of Inflated Expectations' in the year 2015. That review mentions also that the 2015 peak may actually be the second peak of inflated expectations of an "extended hype cycle", were the first peak of this kind occurred already about two decades earlier (Huang 2015).

## 4. Current methodologies and best practices of college-level 'Classroom 3D Printing' in the STEM field

### 4.1. Need for 3D print files and typical print-out times

Before any college educator can print a model in 3D, she or he has to get hold of a suitable 3D print file. This can be achieved by several routes as outlined below. The 3D print file is then to be sent to a 3D printer, which may be provided as part of the college communication/computing/printing infrastructure (Nadel 2017), may be located at a professional print shop, or might even be of the hobby variety and owned privately. The educator should, however, be aware that the actual printing process of a typical model for STEM education may currently take several hours in dependence of the model's complexity. Also, some post processing to remove support material from the 3D print is often necessary.

### 4.2. Open access sources for crystallographic information framework files (CIFs)

Because 'Classroom 3D printing' is already quite mature, one may create 3D print files of small molecule and crystal structure models with atomic level details directly from their crystallographic records in large open-access databases such as the Crystallography Open Database (crystallography.net, Gražulis 2009, 2011, 2016, Bruno 2017). The approximately 400,000 records of this database are in the form of crystallographic information framework files (CIFs), which is the mandated format of the IUCr and computer readable. A human being can discern the information in these files straightforwardly, but may occasionally need to consult the IUCr reference text (Hall 2005), which is also available on-line, or at least the introductory paper (Hall 1991) on this subject.

The entries of the American Mineralogist Crystal Structure Database (AMCSD, Downs, 2003, Bruno 2017) are an integral part of the Crystallography Open Database (COD). The websites of the Open-Access Crystallography project of PSU's Nano-Crystallography group (nanocrystallography.research.pdx.edu) house some 9,000 freely downloadable CIFs with special relevance to college/university education (Moeck 2004, Moeck 2005, and Moeck 2006a,b). There is also a small depository of ready-made 3D print files of small molecules and crystal structures at this website (Moeck 2016b). A mirror of the COD (nanocrystallography.org, Moeck 2006c) is also housed at the servers of PSU's Nano-Crystallography group. This mirror can be assessed over the search interface of the websites of the Open-Access Crystallography project (Moeck 2004).

Note that not all CIFs were *"created equal"* by automatic routes and people alike as the CIF standard is rather complex, comprehensive, and keeps on getting expanded over time. Occasional problems of reading certain CIFs into 3D visualization and 3D print file preparation programs may result from this fact.

Experimentally obtained CIFs of organic molecules and biopolymers often contain information on additional solvent or ligand molecules and counter-ions that are not part of the molecules (Scalfani 2016a). As these additions to the molecule or crystal structure are typically not of interest for the creation of 3D printed models of the molecule or crystal structure, one needs some means to 'clean' a crystallographic information framework file (CIF) up. Several ways of doing so are mentioned below. For the benefit of STEM educators outside of the extended crystallography field, we will discuss below a few core concepts of the CIF format that pertain directly to 3D printing within the STEM education context.

### 4.3. Microsoft Windows^TM executable software for creating 3D print files of individual molecules, crystal morphologies, and longitudinal effect representations of physical properties of crystals

Besides visually appealing atomic-level molecule and crystal structure visualizations, the teaching in the fields of crystallography, condensed matter physics, materials (including nano-materials) science and engineering, and mineralogy/geology requires a range of other visualized objects to help with the understanding of complex concepts in 3D. These other objects that need to be visualized are first and foremost crystal morphologies and representation surfaces of tensors. The latter represent predominantly anisotropic physical properties of single crystals. Tensors are also used to express anisotropies of polycrystalline and non-crystalline materials as discussed within various branches of technology education at the college/university level.

Three separate Microsoft Windows^TM executable programs were developed over the years for each of the above-mentioned types of interactive visualizations, i.e. molecule (and prospectively also crystal) structure (Kaminsky 2014), crystal morphology (Kaminsky 2005, 2007), and physical properties of single crystals (Kaminsky 2000, 2015). In 2014/2015, the latter two programs were extended to include export routines for 3D print files.

While all of this software is completely free to use for non-commercial purposes, maintenance and further developments, i.e. implementation of new features as suggested by users, can only proceed if users (who can afford to do so) donate via the official donation portal of the University of Washington (http://128.95.152.162/donations/donations.htm). For software download links, see below.

The crystal morphologies and tensor representation surfaces can be easily rendered in single, closed meshing (STL) files. The representation of a ball-and-stick model of a molecule involves, however, interleaved cylinders and spheres which results in a multi-mesh representation. Slicing programs can be configured to deal with multi-meshing files, as long as all individual spheres and cylinders are closed and the model is monochromatic. For colored printing of ball-and-stick models of molecules, the VRML file format was found to cause no problems.

We observed in 2014 that the pathways from a CIF, which may contain crystal morphology data, to a 3D printed model of a molecule or crystal structure involved always repetitive steps and information/file transfers from one program to another. We therefore, decided to simplify the whole procedure into a one-click process in the newly developed Cif2VRML program (http://128.95.152.162/Cif2VRMLHome/Cif2VRML.htm, Kaminsky 2014). This program allows for the import of a CIF and directly outputs a STL meshing file for monochrome printing or a VRML file for color prints.

Currently under development is the 'packing of a molecule' into a crystal structure so that 3D print files of the latter type of models can in the future be prepared with this program as well. (As already mentioned above, donations would be very welcome to finish the implementation of this particular task.)

One very useful feature of the Cif2VRML.exe is its capability to load and understand different CIF-file formats as demonstrated by the following examples. The molecular structure of ecstasy was, for example, given in the 'old,' CIF format, that was used before 1997. Figure 3 shows a ball-and stick rendering of this molecule and 3D models that were printed out with both an inexpensive hobby printer and a 3D Systems ColorJet printer. (For technical details on contemporary 3D printing technologies, consult the appendix.) Hydrogen bonds were included in the visualization and 3D printed models by connecting the chlorine ion to the rest of the organic molecule in Cif2VRML.

The molecular structure of sucrose was, on the other hand, available in the current CIF format. In the chosen ball-and-stick rendering of the sucrose model in Fig. 4, the covalent radii option of Cif2VRML was used for all atoms. Both a 3D Systems ColorJet and a MultiJet printer were utilized for the printing of the molecule models.

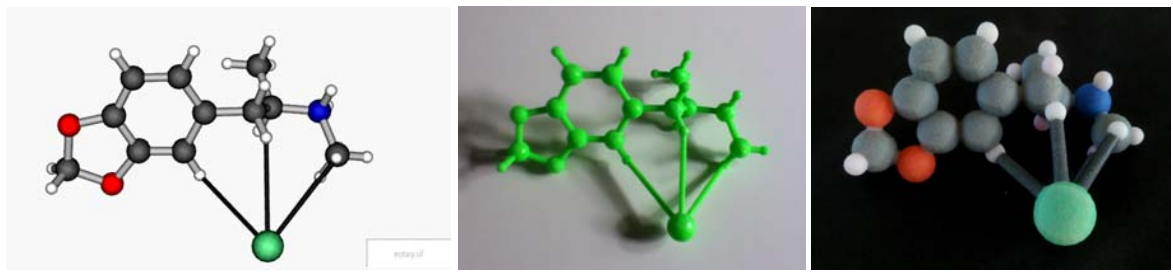

**Figure 3:** Visualizations of an ecstasy molecule, **left** in Cif2VRML, **center** as 3D print obtained with an inexpensive Prusa i2 hobby printer, and **right** with a 3D Systems ColorJet printer. The 3D models measure several centimeters across.

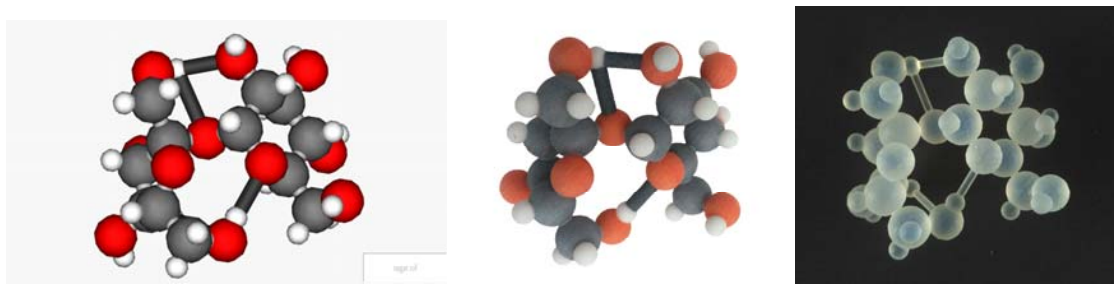

**Figure 4:** Visualizations of a sucrose molecule, **left** in Cif2VRML, **center** as printed out with a 3D Systems ColorJet printer, and **right** as printed out with a 3D Systems MultiJet printer. Note that hydrogen bonds stabilize the real molecule and their representations stabilize the 3D printed models. The 3D models measure several centimeters across.

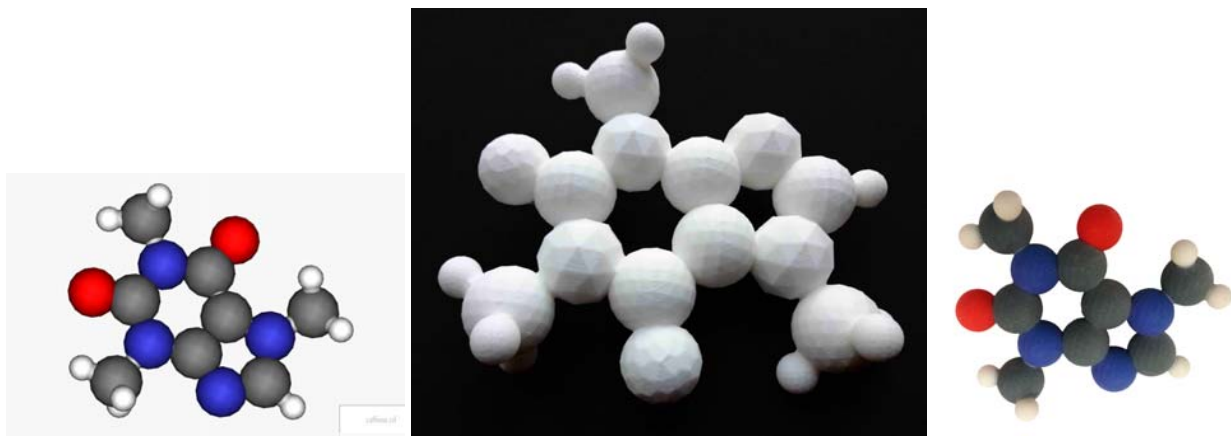

**Figure 5:** Visualizations of a caffeine molecule, **left** in Cif2VRML, **center** as printed out with a 3D Systems SLS Nylon printer (and flipped over to expose the "back side"), and **right** as printed out with a 3D Systems ColorJet printer. The 3D models measure several centimeters across. (The special surface finish of the individual "atoms" in the model in the center is present on both sides.)

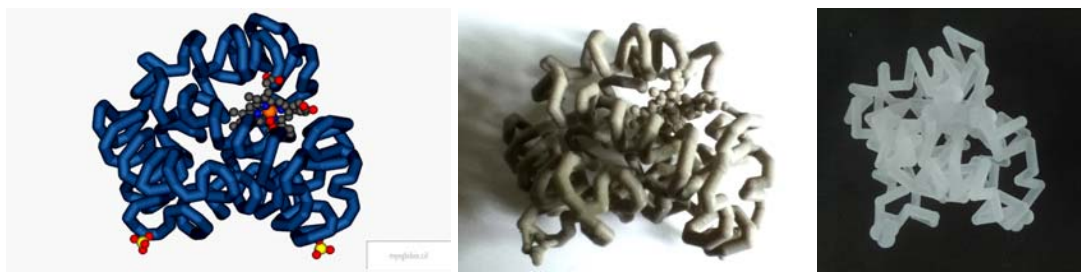

**Figure 6:** Visualizations of a myoglobin molecule, **left** in Cif2VRML, **center** as 3D print obtained with an inexpensive Prusa i2 hobby printer, and **right** as 3D print obtained with a 3D Systems MultiJet printer. The 3D models measure several centimeters across.

The molecular structure of caffeine was given in the mmCIF format. Covalent radii were used again for the 3D rendering. Note the significant differences in the surface finish of the different atom models in the 3D print in the center of Fig. 5 as created with the Cif2VRML program. These kinds of distinct surface finishes allow visually impaired people to differentiate between the most important similar-in-size elements in an organic molecule. A 3D Systems SLS Nylon printer was used for the creation of that particular model, which has exceptional mechanical properties due to both the sturdy design and the selection of nylon as print material. A 3D model of caffeine was also printed out with a 3D Systems ColorJet printer, right-hand side of Fig. 5, to allow for the distinction of the atom types within the caffeine molecule by different colors.

The structure of the protein myoglobin was given in the current CIF format of the Worldwide Protein Data Bank (Berman 2014). Figure 6 shows the structure of this molecule in the Cif2VRML program and as 3D models as obtained with both an inexpensive hobby printer (center) and a 3D Systems MultiJet printer (right).

Geometrically speaking, a crystal morphology is determined by corresponding face (Miller) indices and central distances, i.e. radius vectors from the center of the crystal outwards that are oriented perpendicular to the crystal faces. In addition, only a 'unique' set of faces is needed to complete a crystal's morphology via its 3D point group symmetry. One should note, however, that without deliberate modifications to the central distances of all faces, a crystal morphology model looks unrealistic. (It may even be impossible to recognize the type of mineral from its 3D printed model if one does not define suitable central distances.) For quartz, a randomly applied difference between the central distances of sets of faces that are not symmetry related yields a recognizable model, Figs. 7 and 8.

The WinXMorph program (http://128.95.152.162/WinXMorphHome/WinXMorph.htm, Kaminsky 2005, 2007, 2014) was used for the creation of the 3D print file of the model on the right-hand side of Fig. 7. This software comes with many features including growth sector coloring, simulated cuts through models, and twin simulations. The program is capable of exporting printable STL and VRML files. Figure 7 shows a model of an α-quartz contact twin in the WinXMorph software and as 3D print as obtained with a 3D Systems MultiJet printer.

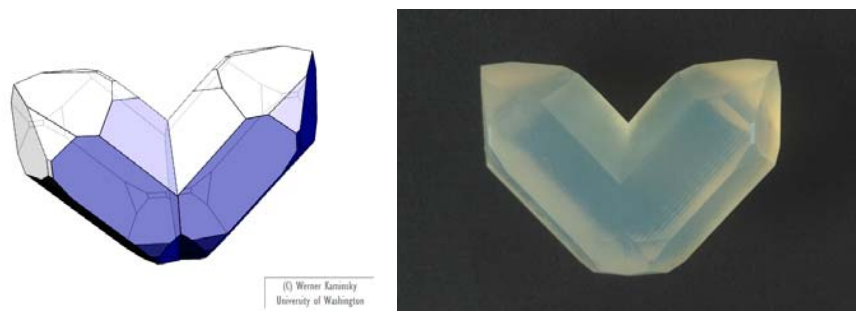

**Figure 7:** Visualizations of a quartz contact twin, i.e. two right-α-quartz crystals grown in contact on the (112) face. **Left** in WinXMorph, and **right** as 3D print model obtained with a 3D Systems MultiJet printer. The 3D model measures several centimeters across and resembles a natural quartz twin rather well due to the chosen print material and technique.

As Fig. 8 demonstrates, the CIF syntax provides a straightforward means to combine crystal morphology with crystal structural data. WinXMorph includes, therefore, import and export routines to handle CIF-file morphology data. The ASCI text excerpt from a CIF on the right-hand side of Fig. 8 shows an example for dextrorotatory (Glazer 2018) right-α-quartz, space group $P3_221$, which is visualized on the left-hand side of Fig. 8. (To simulate the morphology of a left-α-quartz, space group $P3_121$, one only needs to replace face (511) with (141) and exchange central distances for the (011) and (101) faces. For the structures of right- and left-α-quartz, their "structural handedness", and how that translates into opposite rotations of the plane of polarized light, see Glazer 2018.)

The six lines at the top of the list on the right-hand side of Fig. 8 are typically part of an ordinary crystal structure CIF and describe the metric of the unit cell. The magnitudes of the unit cell vectors are given in Å (= 0.1 nm) and the unit cell angles are given in degrees. The line _symmetry_point_group_name_H-M '32' in this list specifies the point symmetry group of the crystal, which is the same for right- and left-α-quartz. This kind of information is often omitted in crystal structure CIFs as it can be straightforwardly recovered from the entry for the space group symmetry of the crystal, which is almost always present.

The following ten lines of the list on the right-hand side of Fig. 8 specify the morphology of the crystal (in connection with the already specified point group symmetry). The above-mentioned Miller indices, e.g. (111), and central distances, e.g. 0.8500, are within a loop that the WinXMorph program interprets on the basis of the provided information on the crystal's point symmetry group, i.e. the symbol '32'. When the crystal's point symmetry group is not specified within a CIF, Miller indices and central distances of all faces for the given morphology need to be included in the CIF-syntax loop (that specifies the crystal morphology), see right-hand side of Fig. 8, so that WinXMorph can visualize the morphology of the crystal and export it to the corresponding 3D print files.

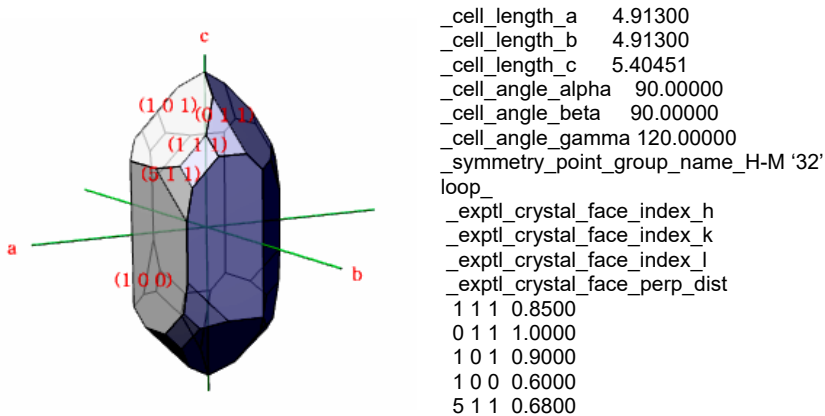

```
_cell_length_a            4.91300
_cell_length_b            4.91300
_cell_length_c            5.40451
_cell_angle_alpha         90.00000
_cell_angle_beta          90.00000
_cell_angle_gamma         120.00000
_symmetry_point_group_name_H-M '32'
loop_
_exptl_crystal_face_index_h
_exptl_crystal_face_index_k
_exptl_crystal_face_index_l
_exptl_crystal_face_perp_dist
 1 1 1  0.8500
 0 1 1  1.0000
 1 0 1  0.9000
 1 0 0  0.6000
 5 1 1  0.6800
```

**Figure 8: Left:** Visualizations of a dextrorotatory right-α-quartz single crystal in WinXMorph. **Right:** ASCI text excerpt from the CIF that was used to create the visualization on the left.

Physical properties of crystals are governed by tensors of different ranks which can be represented by different types of matrices (Schwarzenbach 1997). The rank of a tensor and type of matrix depend on the individual physical property. The nature of the coefficients of the matrices, i.e. which tensor components are non-zero and which are mathematically related to others is governed by the point symmetry group of the crystal as well as the specific symmetry of the physical property itself. This feature leads to rather complex looking visualization of (longitudinal effect) tensor representations which depend typically only on a few individual numbers for cases of highly symmetric crystals.

For example, the optical rotation in low-quartz in Fig. 9 is calculated from just two numbers: the optical rotation along the three-fold axis, and one perpendicular to it, whereby the latter number carries the opposite sign. A color visualization of this tensor surface gives an intuitive representation of the pronounced anisotropy of this physical property, left and center of Fig. 9, as it shows how the magnitude of the optical rotation effect decreases in alpha-quartz for directions progressively off the (±) optic axis/z-axis (red lobes), becomes zero, and then switches its sign (green ring).

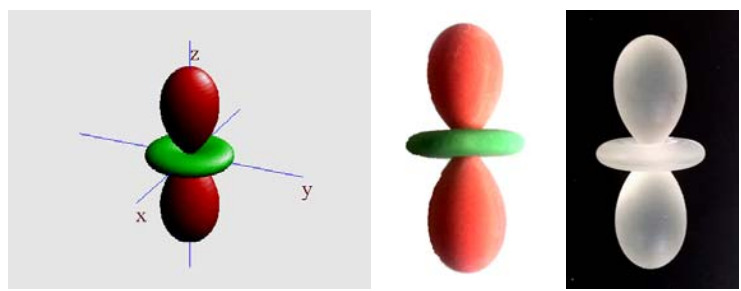

**Figure 9:** Visualization of the optical rotation of α-quartz, **left** in WinTensor, **center** as printed with a 3D Systems ColorJet printer, and **right** as obtained with a 3D Systems MultiJet printer. The 3D models measure several centimeters across. The color model in the center can be used for both right and left α-quartz as one is free to assign opposite signs to the two colors in two different ways.

Dextrorotatory α-quartz will rotate the plane of polarization of light that moves towards an observer in a clock-wise and laevorotatory α-quartz will rotate it in an anti-clockwise manner (Glazer 2018). It is up to the user of the 3D printed model in the center of Fig. 9 to assign the red and green colors to these two senses of rotations in order to represent the (anisotropic) optical-rotation property of either right- or left-α-quartz.

Per definition, the symmetric part of a tensor defines the magnitude of a physical property along all radius vectors that originate at the center of a body that possesses a (longitudinal effect) tensor representation surface (Schwarzenbach 1997). An isotropic physical property such as the refractive index of a cubic crystal is, accordingly, defined by the surface of a sphere.

The software WinTensor (http://128.95.152.162/WinTensorhome/WinTensor.htm, Kaminsky 2000) was expanded to handle physical property/fields/medium symmetries and exports STL as well as VRML renderings of the (longitudinal effect) tensor representation surface of a physical property of a single crystal (Kaminsky 2015) and used to create the 3D print files (in both the VRML and STL format) for the two 3D prints in the center and on the right-hand side of Fig. 9.

*4.4. Open access resources for the creation of 3D print files for bioscientific and biomedical applications as well as of longitudinal effect representations of tensor properties of crystals*

If one is content with handing the interactive creation of a 3D print file for a protein molecule model over to a remotely housed computer program that can be freely accessed, one may use the National Institute of Health's 'NIH print exchange' website (3dprint.nih.gov). This website utilized scripts of the computer program UCSF Chimera (Pettersen 2004). A protocol for obtaining the desired 3D print files by this route is given in the paper by Da Veiga Beltrame 2017. Several options are offered for both the interactive visualization at this website and the creation of the 3D print files.

At that website, one only needs to enter the so-called biopolymer accession code, which one can obtain freely from the websites of the Worldwide Protein Data Bank (wwPDB, Berman 2014, Bruno 2017). The paper by Da Veiga Beltrame 2017 provides also an off-line protocol for the creation of 3D print files of protein molecules with the stand alone UCSF Chimera (Pettersen 2004) software. The 'NIH print exchange' websites feature also a large depository of ready-made 3D print files for bioscientific and biomedical applications (Coakley 2014) so that one may find the 3D print file that one is looking for right there.

The websites of the open-access Material Properties Open Database (MPOD, Fuentes-Cobas 2017a) provide, on the other hand, a large source of 3D print files for the (longitudinal effect) representation surfaces of tensor properties of crystals (Fuentes-Cobas 2017b). Again, one is handing the creation of the 3D print files over to a remotely housed computer program that can be freely accessed. There are several options for the interactive visualization, but only *.stl files with low and medium resolutions can currently be downloaded. (STL files can become very large in dependence of the resolution and complexity of a model so that restrictions on model resolutions are necessary in order to obtain reasonable downloading times.)

*4.5. Interactive creation of 3D print files for atomic-level crystal structure and individual molecule models at a dedicated website that is provided by PSU's Nano-Crystallography group*

While current methodologies and best practices of the creation of 3D print files for small and large molecules (including biopolymers) have been discussed in the preceding two sections, the reader of this book chapter has so far not been advised on how to create 3D print files of atomic-level crystal structure models effectively. As it is well known, single crystal X-ray crystallography (Schwarzenbach 1997) delivers the structure of a molecule as it is "packed" into a crystal structure. The individual molecule possesses a 3D point group symmetry in isolation, while the coordinates of the atoms in the same molecule are also characterized by a 3D space group symmetry when it is packed into a crystal structure.

The situation is somewhat different for inorganic crystals including minerals where there are no individual molecules in a crystal structure. When referring to an inorganic crystal structure, a CIF that was derived by means of single crystal X-ray crystallography, contains, therefore, no information on individual molecules and how they are packed. Such a CIF is for example given in Fig. 10 for a specimen of the mineral right-α-quartz.

Note that the coordinates of only one silicon atom and one oxygen atom, which form together the asymmetric unit, is mentioned in this CIF, Fig. 10, as it also contains information on the space group symmetry of the mineral, i.e. the symbol 'P 32 2 1' at the end of the line '_symmetry_space_group_name_H-M'. If one were to use some standard CAD program to create a 3D print file of a crystal structure model, one would almost certainly have to specify the positions of all atoms in the model. When one is utilizing the interactive route to the creation of 3D print files from CIFs that we describe below in this section, on the other hand, atomic position information in the unit cell in its standard form suffices as shown in Fig. 10. (The CIF in Fig. 10 contains also experimentally obtained anisotropic atomic vibration parameters, but this information is currently not utilized by us for the creation of 3D print files.)

This particular CIF was utilized for the creation of the 3D print file from which the 3D model in Fig. 11 was printed and is openly accessible at the COD, its mirror at PSU, the AMCSD, and the Open-Access Crystallography project that provides also access to the local COD mirror, see Fig. 12, in addition to collections of CIFs with special relevance to STEM education.

```
#---------------------------------------------              _diffrn_ambient_pressure          100
#$Date: 2016-02-16 14:49:47 +0200 (Tue, 16 Feb 2016) $      _exptl_crystal_density_diffrn   2.646
#$Revision: 176465 $                                        _cod_original_formula_sum       'Si O2'
#$URL: svn://www.crystallography.net/cod/cif/9/00/07/9000775.cif $   _cod_database_code        9000775
#---------------------------------------------              loop_
#                                                           _symmetry_equiv_pos_as_xyz
# This file is available in the Crystallography Open Database (COD),   x,y,z
# http://www.crystallography.net/. The original data for this entry   y,x,2/3-z
# were provided the American Mineralogist Crystal Structure Database,  -y,x-y,2/3+z
# http://rruff.geo.arizona.edu/AMS/amcsd.php                          -x,-x+y,1/3-z
#                                                                      -x+y,-x,1/3+z
# The file may be used within the scientific community so long as      x-y,-y,-z
# proper attribution is given to the journal article from which the   loop_
# data were obtained.                                                 _atom_site_aniso_label
#                                                                     _atom_site_aniso_U_11
data_9000775                                                          _atom_site_aniso_U_22
loop_                                                                 _atom_site_aniso_U_33
_publ_author_name                                                     _atom_site_aniso_U_12
'Levien, L.'                                                          _atom_site_aniso_U_13
'Prewitt, C. T.'                                                      _atom_site_aniso_U_23
'Weidner, D. J.'                                                      Si 0.00854 0.00716 0.00725 0.00358 -0.00001 -0.00002
_publ_section_title                                                   O 0.01745 0.01322 0.01229 0.00973 -0.00291 -0.00408
;                                                                     loop_
  Structure and elastic properties of quartz at pressure              _atom_site_label
  P = 1 atm                                                           _atom_site_fract_x
;                                                                     _atom_site_fract_y
_journal_name_full                'American Mineralogist'             _atom_site_fract_z
_journal_page_first               920                                 Si 0.46970 0.00000 0.00000
_journal_page_last                930                                 O 0.41350 0.26690 0.11910
_journal_volume                   65
_journal_year                     1980
_chemical_formula_sum             'O2 Si'
_chemical_name_mineral            Quartz
_symmetry_space_group_name_H-M    'P 32 2 1'
_cell_angle_alpha                 90
_cell_angle_beta                  90
_cell_angle_gamma                 120
_cell_length_a                    4.916
_cell_length_b                    4.916
_cell_length_c                    5.4054
_cell_volume                      113.131
```

**Figure 10:** CIF of dextrorotatory crystalline right-α-quartz from which the 3D print file was created that led to the printed model in Fig. 11. The asymmetric unit consists of two atoms for which the fractional coordinates are listed in the last two lines of this CIF. The (chemical) formula unit of this mineral, on the other hand, consists of one silicon atom and two oxygen atoms, i.e. $SiO_2$.

As mentioned above, the CIF in Fig. 10 contains *all* off the crystallographic information that was needed to create the X3D print file that was utilized in the printed-out 3D model of the 'internal structure' of a right-α-quartz crystal in Fig. 11, which may have an 'external morphology' close to what is shown on the left-hand side of Fig. 8. While X3D is a XML version of VRML that contains color information, printing in color was not necessary as there is an unmistakable difference in the sizes of the printed silicon and oxygen atom representations in Fig. 11 as automatically chosen by our 3D converter tool (Moeck 2017a,b), see Fig. 13.

Figure 12 shows a screen shot of the significantly abbreviated search for the CIF of Fig. 10 at the local PSU mirror of the COD under the search interface of the Open-Access Crystallography project (Moeck 2004). (Such searches can be done either independently or directly from the PSU 3D Converter.) The full search for the elements 'Si' and 'O' as combined to form 'quartz' with the option "Strict number of elements = 2" yielded on May 13, 2017, fifty returns, some of which from studies under elevated pressures and temperatures. (Without restricting the search to two elements, we obtained 77 results on that day. Note in passing that the CIF in Fig. 10 refers to experimental measurements at approximately 1 atmosphere pressure (100 kPa).)

The development of our Internet-based tool for the interactive creation and export of 3D print files of crystal and molecule structures from CIFs, the 'PSU 3D Converter' – Fig. 13, was inspired by the recent work of Bob Hanson in collaboration with Vincent Scalfani and co-workers (Scalfani 2016a). Our tool has been freely accessible over the websites of the Open-Access Crystallography project (http://nanocrystallography.research.pdx.edu/3d-print-files/ convert, Moeck 2017a) since the spring of the year 2017.

The middle of the upper frame of Fig. 13 features a small fragment of the structure of right-α-quartz with the visualization of a silicon atom (large light-brown ball) at the center. While this atom is displayed as bonded to four (smaller) oxygen atoms, the inherent space group symmetry of the low-quartz mineral allows only for a distorted tetrahedral coordination, which is sufficiently small to not readily be detectable in Fig. 13.

We display the right-α-quartz structure fragment in this figure as so called 'unpacked' unit cell in the 'Axes off' option. There are somewhat subtle differences between 'unpacked' and 'packed' unit cells for right-α-quartz. In our particular case, the 'Packed Cell' button adds two Si atoms to the JSmol display that are shared between four unit cells each. 'Packed Cells' show always the full content of a unit cell but also overhangs of atoms into neighboring unit cells. (Other crystal structures may feature more dramatic differences between 'unpacked' and 'packed' unit cells.)

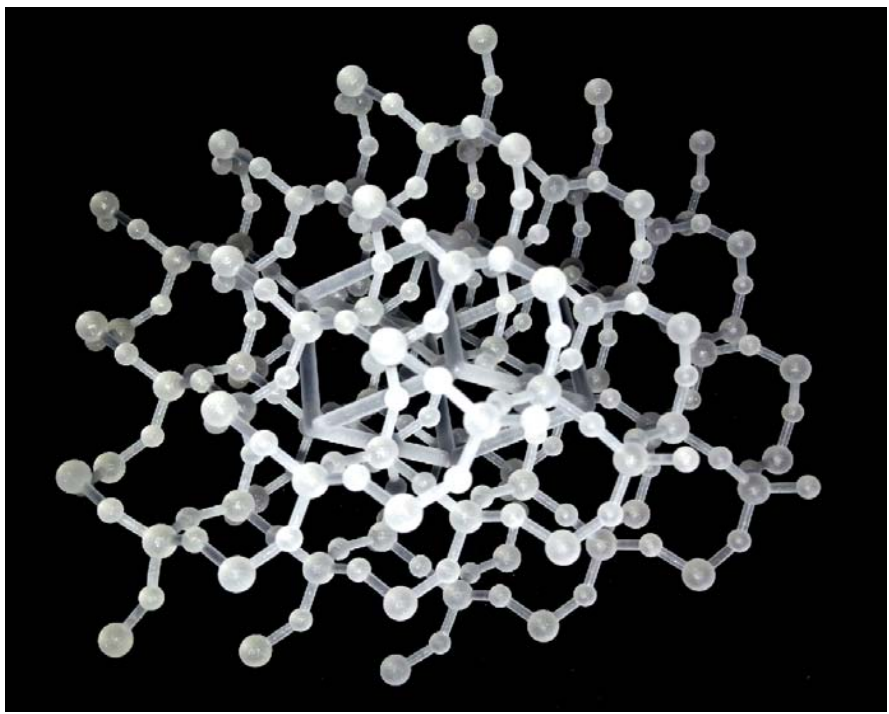

**Figure 11:** 3D printed ball-and-stick model of right-α-quartz for which the print file was created from the CIF in Fig. 10 with the Internet-based PSU 3D Converter shown in Fig. 13. A 3D Systems ProJet MJP 2500 printer and highly flexible "engineering Armor" material was utilized. Note that this particular 3D printed model is highly durable and fracture resistant in spite of the numerous small features. The 3D print file was exported while the unit cell outlay option ('Axes' in Fig. 13) of our converter tool was switched to 'on'. The 3D model measures approximately 20 cm along its largest dimension.

Formula:- O2 Si -
**Cell Parameters:** $a = 4.916$Å $b = 4.916$Å $c = 5.4054$Å $\alpha = 90.0°$ $\beta = 90.0°$ $\gamma = 120.0°$
**Cell Volume** = 113.131Å$^3$
View CIF        Download CIF        View Structure in JSmol with HTML 5
View Structure in Jmol V12 with Java

**Figure 12:** Screenshot of a very small section of a search for 'quartz' at the local mirror of the COD over the websites of the Open-Access Crystallography project. Note that there are links for viewing, downloading, and visualizing the content of the CIF. We recommend visualizing the CIF with JSmol (with HTML 5) as that does not require that Java is installed (and properly enabled) at the local machine. The chemical formula, crystal unit cell parameters, and unit cell volume of the searched entry are displayed.

While the upper frame of Fig. 13 (and of the corresponding web-page) contains the JSmol based 3D modeling and viewing interface, the lower frame represents an abbreviated section of the search interface of the Open-Access Crystallography project with the default setting to the local mirror of the COD. In the following, we walk the reader through the process of creating a 3D print file from a CIF at the 'PSU 3D Converter' website. While doing so, we follow the 'number labels' in Fig. 13 for easy reference. For simplicity, we stick to an atomic-level structure model of right-α-quartz.

1. First, the user can search the local COD mirror for quartz in the lower frame of the website, Fig. 13, using the constituent elements, 'O' and 'Si', and/or the key word 'quartz', and/or '2' for the 'Strict number of elements' as separate or combined search categories. Users can also choose elements to search for in chemical compounds from the displayed periodic table of elements. When the search criteria are specified, one needs to press the ENTER key or click the 'Search' button.

2. Next, the lower frame of the website (of the 'PSU 3D Converter') will show at its bottom all matching entries form the local COD mirror as shown in largely abbreviated form in Fig. 12. The user then chooses a specific CIF there and clicks "Download CIF". This redirects one to the upper frame of our 3D converter tool (without actually downloading the CIF to the local machine of the user when done within our converter tool).

3. In the upper frame of the website ('PSU 3D Converter'), the user then clicks the 'Load File' button and follows the menu to upload the CIF that was selected in the previous step. Our converter tool will then show a 3D visualization of a single 'packed' unit cell of the crystal structure by default. It is also possible to load CIFs that one has obtained by other routes (including individual files from a range of curated crystallographic databases, Bruno 2017) because pressing the 'Load CIF' button opens an upload menu.

4. Now, one may create a 2 nm by 2 nm by 2 nm packed right-α-quartz crystal structure model block by means of JSmol by clicking the '20-Angstrom Cube' button of our converter. The left-hand side of Fig. 14 shows the "structure cube" just created. Such a cube contains approximately 70.7 unit cells of this mineral and appears at the center of the upper frame of the 'PSU 3D Converter' next. A consequence of displaying this cube for this particular crystal structure is that a bunch of oxygen atoms at the upper left edge appear to be un-bonded. They can be straightforwardly removed from both the display and exported 3D print files by pressing the 'Largest Molecule' button in our 3D converter tool.

5. Next one may press the 'Axes On/Off' button to toggle the display of the unit cell outline 'on' or 'off'. If left displayed, the outline of one unit cell will be included in the 3D print file to be exported in the next step. Since a length scale for the printed-out model is already set by the '20-Angstrom Cube' option, one does not really need to set an additional length scale by displaying and printing the unit cell outline in addition.

6. Finally, one has to click the 'Save X3D button' (or 'Save STL' or 'Save VRML' button) to export a 3D print file of the visualized model in the chosen 3D print file format. One has to enter the name of the file to which the exported data is to be saved. The user's browser may prompt for additional confirmation before saving the just created 3D print file to the local computer of the user.

One may modify steps 4 to 6 by going for an atomistic model of the right-α-quartz crystal structure that comprises nine unit cells as shown on the right-hand side of Fig. 14 and in Fig. 11. The central unit cell outline in the JSmol visualization comes in very handy here for setting a length scale to the printed model, i.e. $a = b$ will be equivalent to approximately 0.49 nm and $c$ will be equivalent to approximately 0.54 nm (as shown in Figs. 10 and 12).

The '3x3x3 Cube' button of the 'PSU 3D Converter', Fig. 13, displays/produces a genuine cube only for crystal structures with cubic Bravais lattice types. This button 'packs' the content of nine unit cells into a single block, which is very useful for creating 3D ball-and-stick models that show a quite representative part of an inorganic crystal structure, see Fig. 11. A JSmol visualization of such a block is shown on the right-hand side of Fig. 14.

While all oxygen atoms are in this visualization at positions within unit cells, four of the silicon atoms per unit cell are at positions on unit cell edges and two silicon atoms per unit cell are at positions at unit cell faces. The representations of atoms at unit cell edges are in the 3D printed model intersected by cylinders that represent the unit cell outline, Fig. 11. As all atoms are also connected to other atoms in the 3D printed model, the cylinders of the unit cell outline provide extra stability to the printed model. Atoms at positions on unit cell edges contribute one quarter of the atom to an individual unit cell and atoms at positions on unit cell faces contribute one half of the atom to that unit cell. (In cases where no atom possesses a zero in at least one of its fractional coordinates in the unit cell as provided in the CIF, the cylinders of the unit cell outline will for obvious reasons not be connected to the rest of the crystal structure in the 3D printed model so that they cannot contribute to the mechanical stability of the model.)

As already mentioned above, printing a unit cell outline into an atomic-level crystal structure model is very helpful for conveying a sense of the length scale of the model as the magnitudes of the basis vectors are given in Å in the CIF from which the 3D print file was created. The '20-Angstrom Cube' button of the 'PSU 3D Converter' does, on the other hand, always display/produce a genuine cube of 8 $nm^3$ into which a crystal structure is fit into, see left-hand side of Fig. 14.

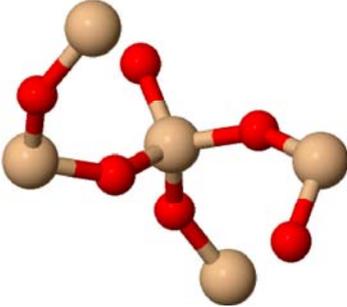

**Figure 13:** Screenshot of the Internet-based 'PSU 3D Converter' (Moeck 2017a,b) with the content of one 'unpacked' unit cell of right-α-quartz displayed in the JSmol visualization and editing workspace in the middle of the upper frame under the 'Axes off' option.

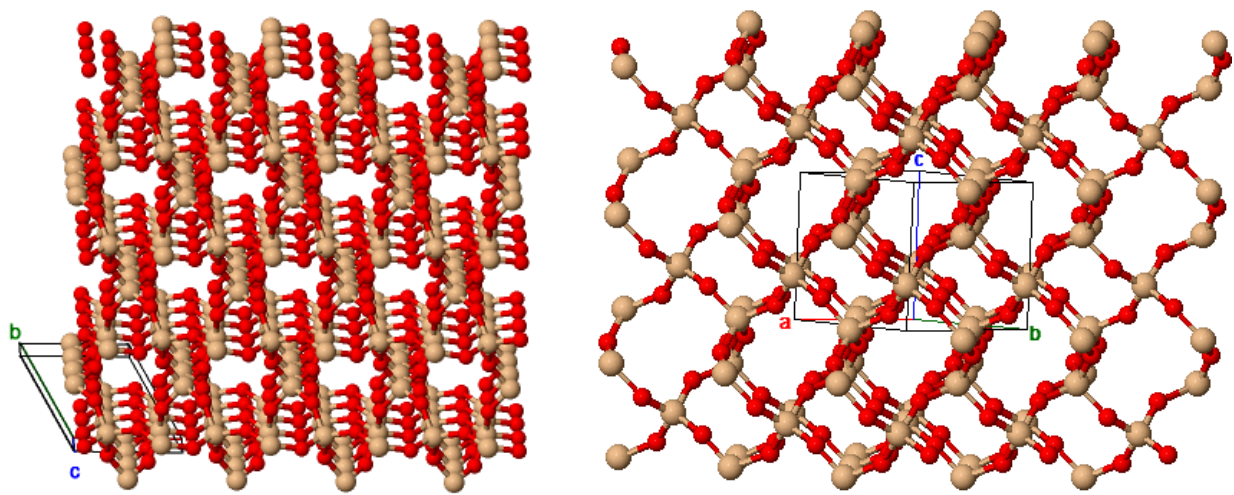

**Figure 14: Left:** JSmol display of a cube with 20 Å edges (8 nm³ volume) containing the atomistic structure of right-α-quartz. **Right:** JSmol display of the content of nine (i.e. 3 x 3 x 3 packed) unit cells of right-α-quartz.

Thanks to Bob Hanson's tireless work (Hanson 2010) over many years, JSmol (as incorporated into our 3D converter tool) provides a very wide range of additional 3D model editing and displaying capabilities. To access them, one first right-clicks anywhere in the JSmol visualization/editing space of the 'PSU 3D Converter' to reveal the contextual menus. Whatever one may choose to display as a result of JSmol editing in the center of the upper frame of the website (of the 'PSU 3D Converter') will be faithfully exported to the 3D print file in the chosen format.

For example, if one does not use the shortcut procedure executed by the 'Largest Molecule' button in Fig. 13, one may select 'Set Picking' > 'modelKitMode' in the applicable JSmol sub-menu, right-click on the model viewer space of our converter tool and select 'delete atom'. Next, one needs to left-click on individual atoms to actually delete them from both the interactive visualization and 3D print files that are to be exported. This procedure does, however, not delete the related crystallographic information in the read-in CIF. In the end, one needs to exit 'modelKitMode' by opening the right-click menu and choosing '...' > 'exit modelKitMode'.

Solvent and ligand molecules as well as counter-ions may be part of an experimental CIF of a small or protein molecule (Scalfani 2016a) at mentioned above in section 4.2, but are typically not present in inorganic crystal structures. (Crystal water is, on the other hand, a well-known integral part of certain mineral structures.) Note that our converter tool allows often for the suppression/removal of typical solvent molecules and counter-ions from the interactive display at the website and from the produced 3D print file simply by the pressing of the 'Largest Molecule' button. That procedure does, however, not work for mechanically interlocked molecules such as the one discussed below, because these bonds represent a recent addition to supra-molecular chemistry, see Sir Fraser Stoddart's Nobel Prize lecture (Stoddart 2017).

Since the structure of right-α-quartz is rather simple with only three atoms in the chemical formula unit, we tested the 'PSU 3D Converter' also with the much more sophisticated structure of a doubly interlocked [2]catenane molecule, Fig. 15. The chemical (sum) formula of that molecule (including solvents and counter-ions) was experimentally derived to be $C_{130}H_{144}F_{36}N_{16}O_{36}P_4Pd_2S_4$ (Peinador 2009). Such molecules have been referred to as Solomon Knots/Links in the popular culture after King Solomon of the Old Testament. Only a few mechanically interlocked catenane molecules have so far been synthesized, see Sir Fraser Stoddart's Nobel lecture (Stoddart 2017).

The experimentally determined atomic structure of that [2]catenane molecule including solvent molecules and counter-ions is displayed in JSmol on the left-hand side of Fig. 15. The experimentally derived CIF of this molecule was obtained from the supporting material of the paper by Peinador 2009 and read into the 'PSU 3D Converter'. Our converter tool was then used to prepare a X3D file for this molecule, which was sent as email attachment to 3D Systems Corporation, where it was printed out, see center of Fig. 15.

The removal of (unconnected) solvent molecules and counter-ions could for this doubly interlocked molecule not be obtained by clicking on the 'Largest Molecule' button of the 'PSU 3D Converter' so that we removed them with JSmol editing functions that are accessible over the 'Set Picking' menu and its sub-menus (as described above). When one presses the 'Largest Molecule' button in our converter tool, Fig. 13, with the CIF of the doubly interlocked [2]catenane molecule read in, one obtains only the "separated molecule component" in the lower-right part of Fig. 15. The 'molecule by number' button of Bob Hanson's 'Jmol Crystal Symmetry Explorer', Fig. 16, facilitates, on the other hand, the separate visualization of both components of the doubly interlocked molecule, from which separate 3D print files can then be exported.

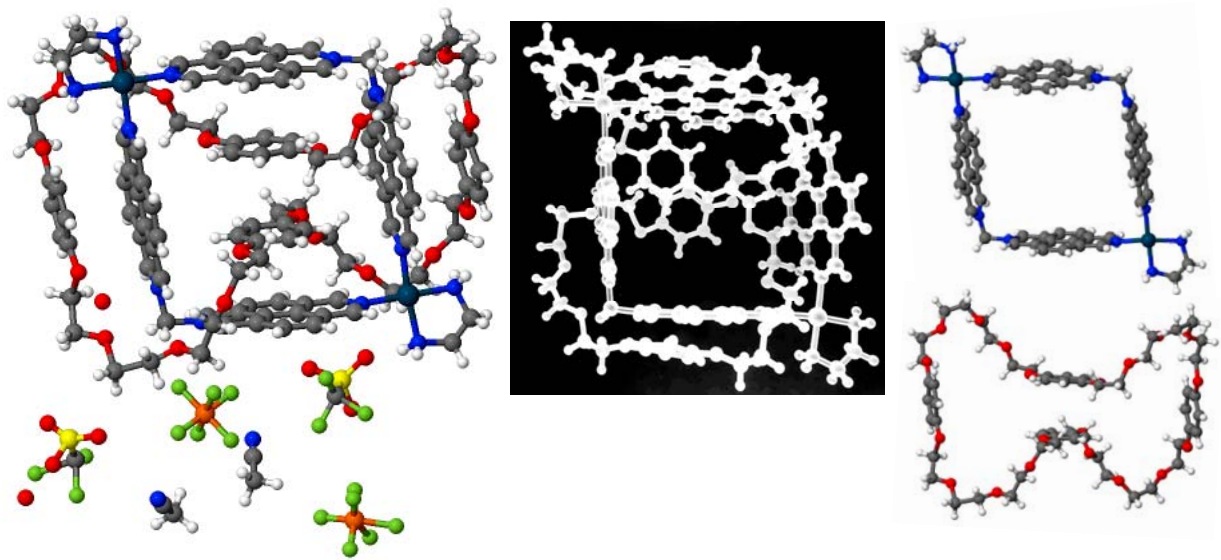

**Figure 15: Left:** JSmol display of the experimentally derived atomic structure of a doubly interlocked [2]catenane molecule (including counter-ions and solvent molecules) as present in a CIF that is freely available (Peinador 2009). **Center:** 3D printed ball-and-stick model of the doubly interlocked molecule as obtained with a 3D Systems ProJet MJP 2500 printer in the highly flexible "engineering Armor" material. Note that the model is highly durable and fracture resistant even with the numerous small features. The model measures approximately 25 cm along its largest dimension. **Right:** JSmol display of the two otherwise doubly interlocked components of this molecule in separated form.

*4.6. Other resources around 3D print files of atomic-level small molecule and crystal structure models*

The 'PSU 3D Converter' as discussed in the previous section is based on Bob Hanson's freely available 'Jmol Crystal Symmetry Explorer' (Hanson 2016a), Fig. 16, which is part of his well known Jmol/JSmol suite of interactive atomic structure visualization/editing tools (Hanson 2010). Note that JSmol is the non-Java HTML 5 version of Jmol.

The 'Jmol Crystal Symmetry Explorer', Fig. 16, possesses the same interactive Jmol/JSmol visualization and editing functionality as the 'PSU 3D Converter' but many more ready-made 'shortcut buttons' at the website level. It was also used for the programmatic conversion of more than 30,000 CIFs from the COD (entries between id 1000000-1999999 and 2000000-2999999) into 3D print files of small molecule models (Scalfani 2016a). The scripting capabilities of the Jmol/JSmol suite were utilized for that purpose.

These 3D print files can be downloaded for free from the figshare platform (Scalfani 2016b). It is also possible to access these ready-made 3D print files from a dedicated website (Hanson 2016b) that features a search surface similar to that of the COD. Bob Hanson's 'JSmol 3D print' website (Hanson 2016b) allows also for both searches of the whole COD (from its inbuilt search surface) and the creation of 3D print files from any CIF in the COD or from any user provided CIF (Scalfani 2016a).

Note that files in the X3D format can be exported from the website of the 'JSmol Crystal Symmetry Explorer', while most other routes end currently only with STL or VRML files. While the first of these two file formats is the industry wide standard for monochrome 3D printing, VRML and X3D allow for color prints in 3D as already mentioned above. STL files can become very large for highly sophisticated models and high resolution as mentioned above, but X3D files for the same models are typically only about one tenth of the size of files in the former format and can, therefore, still be sent as email attachments (as we have done ourselves, as mentioned above).

**Figure 16:** Screenshot of the 'Jmol Crystal Symmetry Explorer' at its original location with the content of one 'packed' unit cell of right-α-quartz (projected down the c-axis) displayed in the JSmol visualization/editing space (with the unit cell outline displayed).

Note also that Bob Hanson has incorporated (STL, X3D, and VRML) 3D print file export capabilities also into his Jmol/JSmol releases from version 14.6.4 onwards (Hanson 2016c). Any website that features a Jmol/JSmol workspace based on any of those recent releases allows, thus, for the interactive creation of 3D print files of models with atomic level details. There is, thus, a generic route from crystallographic information in multiple formats (Hanson 2016c) and CIFs to 3D print files of crystal and molecule structures that is independent of the operating system of the user's computer. The Jmol/JSmol featuring websites of the open-access COD, wwPDB, and AMCSD as well as those of the (subscription based) Cambridge Structural Database (CSD, Groom 2016, Bruno 2017) where access to individual small molecule structure data files is unrestricted will all allow for the remote creation of 3D print files from inbuilt Jmol/JSmol workspaces after necessary upgrades (Scalfani 2016a). 'Classroom 3D printing' is, therefore, at the present time probably as mature as it may become thanks to the efforts of numerous educators at the college/university level who volunteered their time over recent years to reach this state of affairs.

The curators of the CSD provide also a route to 3D prints from CIFs using a stand-alone program (Wood 2017). Their highly venerable (but freely useable) molecule and crystal structure visualization program 'Mercury' (Macrae 2008), that is available for Windows, Linux, and MacOS platforms, can be used for this purpose (after upgrades to versions released in 2016 and later). 'Mercury' allows also for the removal of solvent molecules and counter-ions that a CIF might contain from both the CIF and the 3D print files that are created from it. We employed these editing capabilities of 'Mercury' to prepare a "cleaned up" CIF for the doubly interlocked [2]catenane molecule mentioned in the latter part of section 4.5.

That "cleaned up" CIF of this molecule was read into the Cif2VRML program and 3D print files were exported by means of the proverbial 'one click' (Kaminski 2014). We also tested the 3D print file creation process of 'Mercury' on the basis of the "cleaned up" CIF of this molecule. As a result of these tests and our report in the latter part of section 4.5, we can confidently state here that "mechanical bonds" (Stoddart 2017), which exist as a novel type of bonds between different parts of mechanically interlocked molecules, do not present difficulties to our JSmol based 3D converter tool, to the Cif2VRML software, and to the Mercury software. These kinds of bonds should also not present difficulties to any modern 3D printer that uses support material which is to be removed from the final printed-out model.

Note finally in passing that the bulk of the considerable amount of work of putting hundreds of thousands of CIFs (Bruno 2017) into open access and of developing straightforward routes for the creation of 3D print files from them was done by largely unpaid volunteers (as they were mostly University/College faculty members and their students). That kind of work over more than fifteen years did, however, "force" the commercial CSD to add 3D print file creation capabilities to their 'Mercury' program from the summer of 2016 onwards. The CSD staff's assertion that it is their recent work that *"truly democratises the use of 3D printing for experimentally accurate molecular, inter-molecular and supra-molecular models"* (Wood 2017) is, therefore, quite unfounded.

## 5. Summary and conclusions

The literature on 3D printed models for Science, Technology, Engineering, and Mathematics (STEM) education at the college/university level was reviewed. The prediction by the Gartner consulting company that it will take more than 10 years (from July of 2014 onwards) for 'Classroom 3D Printing' to reach its 'Plateau of Productivity' in one of their hallmark 'Hype Cycle' graphs was critically assessed. Current methodologies and best practices of college-level 'Classroom 3D Printing' were described in the main section of this book chapter. Based on the quite mature state of the affairs as described in that section, it is concluded that Gartner's prediction lacks optimism and seriously underestimates the creativity and resourcefulness of college/university educators as well as their commitment to their students.

Microsoft Windows[TM] executable computer programs for the creation of 3D print files of small molecules and proteins, crystal morphologies, and (longitudinal effect) representation surfaces of tensor properties of crystals were reviewed. A straightforward route from crystallographic information framework files (CIFs) at the Open-Access Crystallography project of Portland State University's Nano-Crystallography group to 3D print files for atomic-level crystal and molecule structure models was described in some detail. Only the exported 3D print files are downloaded from the website of the 'PSU 3D Converter' as no local installation of any supporting program or applet is necessary. Openly accessible depositories of 3D print files and a few other creation routes of 3D print files for STEM education at the college level were also mentioned.

## Appendix: Brief 3D Printing Technology Review

There are numerous additive manufacturing technologies commonly used today. Over the past three decades, major advances have been made and numerous new techniques have been developed. Each of these technologies achieves the basic value proposition of 3D printing, but each has also unique capabilities in terms of material properties, part attributes, as well as costs and printing speeds. This book chapter shows examples form a few of the more common technologies used for rapid prototyping. Both "hobby" grade Fused Filament Fabrication (FFF) and "professional grade" 3D printed models, as obtained by the Selective Laser Sintering (SLS[®]), MultiJet Printing (MJP[®]) and ColorJet Printing (CJP[®]) technologies of 3D Systems Corporation are shown.

Each of these technologies operates by dramatically different physical processes. Many consider Charles Hull the inventor of 3D printing. His company, i.e. 3D Systems Corporation, created the very first commercial 3D printer in the late 1980's using the Stereolithography (SLA[®]) process that he had invented. SLA[®] exposes photo-polymers to radiation (which is typically ultraviolet light). The radiation triggers a chemical reaction within the material, causing curing of the polymer. SLA[®] systems print with supports, are advantageous due to the speed and possible size of prints – both large and small, and can rapidly manufacture parts of different geometries at the same time. They are designed to produce both prototypes and end-use parts of versatile sizes and applications. The SLA[®] technology is particularly good for parts requiring optical clarity. SLA® parts are strong enough to be machined and can be used as master patterns for injection molding, thermo-forming, blow-molding and in various metals casting techniques.

The Selective Laser Sintering (SLS®) method fuses powder materials layer by layer until the structure is built. To do this, a layer of material is spread evenly over a bed. Selected sections of this powdered layer are laser-fused by complete or partial melting. SLS® can be used for a wide range of powder materials, including different types of plastics, metals, ceramics, as well as glass, and can produce structures with high geometric complexity. It is also robust to complex overhangs due to an inherent support structures that is created by the powdered bed. SLS® uses traditional material powders as the raw material and is well known to achieve sufficient physical properties for end-use parts similar to those traditionally manufactured by injection molding.

Direct Metal Sintering (DMS®) refers to 3D Systems Corporation's metal printing process. This process is similar to SLS®, but uses metal powders rather than plastics. This technology can be overly expensive for most educational purposes and is used primarily in medical and aerospace applications, where low volumes of unique and complex models are needed.

The Color Jet Printing process (CJP®) uses inkjet technology to deposit a liquid binder across a bed of powder. The powder is released and spread with a roller to form each new layer. CJP® creates large-build prints in spectacular true-to-life color and is, therefore, ideal for educational uses due to its ability to make color models with sufficient material properties mixed with a combination of fast printing speed and low run cost.

The Multijet™ printing process (MJP®) utilizes a high precision 3D inkjet printing process. This ink-jet technology is combined with wax/resin and/or UV curable materials to produce highly detailed and accurate physical prototypes. High resolution is attainable using a support material that can be easily removed by post processing. MultiJet™ printing is an extremely easy to use and versatile technology because the supports are automatically created and are easy to post process. Such printers can utilize materials with a wide range of mechanical properties. New developments have introduced materials with exceptional mechanical stability simulating the performance of acrylonitrile butadiene styrene and polypropylene, e.g. the above mentioned "engineering Armor", thus turning MJP® into a very good printing process that serves the needs of educators.

Extrusion-based printing, i.e. the FFF technique, consists of the deposition of melted thermoplastics in layers. A bed is placed underneath a heated nozzle which then extrudes molten plastic onto the bed. This technology is ideal for hobbyist and consumer printing for educational needs due to the available low-cost machines and materials as well as relatively simple and safe printing and post processing processes. 3D printed surfaces can, however, be somewhat rough when created with the FFF technology and a model may suffer from the visibility of "layer lines". The 3D prints that are obtained with this technology are nevertheless typically good enough for many educational uses.

So many 3D printing technologies exist because of the numerous different valuable attributes needed by many different kinds of paying customers. It is easy to understand that any technology that has withstood marketplace pressures for some time provides, by definition, unique and value-added features and/or capabilities. Until those unique value propositions are mitigated by a single technology, numerous different technologies will remain for the user to choose from.

**Acknowledgements**

Financial support from both Portland State University's Faculty Enhancement program and the US National Committee for Crystallography is acknowledged. Robert van der Meulen of the Gartner consulting company is thanked for both Figure 1 and for communicating Gartner's definition of 'Classroom 3D Printing' to the first author. 3D Systems Corporation is thanked for facilitating the 3D printing of the models in Figures 4, 5, 7, 9, 11, and 15 as well as for donating Cube 3 printers to WK and PM. We are grateful to Maria Kaminsky, who created the 3D printed models that are shown in the centers of Figures 3 and 6 with a Prusa i2 printer after substantial print parameter optimizations. Prof. Bob Hanson of St. Olaf College, Northfield/Minnesota, is thanked for both his great work over many years in connection with the development of Jmol/JSmol and his generosity as demonstrated by him making the results of these developments openly available. The countless contributors to the COD and AMCSD are thanked for their uploading of crystallographic data sets for the greater good.

# References

Bain, G. A., Yi, J., Beikmohamadi, M., et al. 2006. Using Physical Models of Biomolecular Structures to Teach Concepts of Biochemical Structure and Structure Depiction in the Introductory Chemistry Laboratory. *J. Chem. Educ.,* 83: 1322-1324.

Berman, H. M., Kleywegt, G. J., Nakamura, H., et al. 2014. The Protein Data Bank archive as an open data resource. *J. Comput. Aided Mol. Des.* 28: 1009-1014, http://www.wwpdb.org/.

Bharti, N., and Singh, S. 2017. Three-Dimensional (3D) Printers in Libraries: Perspectives and Preliminary Safety Analysis. *J. Chem. Educ.* 94: 879-885.

Blauch, D. N. and Carroll, F. A. 2014. 3D Printers can provide an added dimension for teaching structure – energy relationships. *J. Chem. Educ.* 91: 1254-1256.

Brown, M. L., Van Wieren, K., Tailor, H. N., et al. 2018. Three-dimensional printing of ellipsoidal structures using Mercury, *CrystEngComm* 20: 271-274.

Bruno, I., Gražulis, S., Helliwell, J. R., et al. 2017. Crystallography and Databases, *Data Science Journal* 16: 38, pp. 1-17.

Carroll, F. A. and Blauch, D. N. 2017. 3D Printing of Molecular Models with Calculated Geometries and p Orbital Isosurfaces. *J. Chem. Educ.* 94: 886-891.

Casas, L. and Estop, E. 2015. Virtual and Printed 3D Models for Teaching Crystal Symmetry and Point Groups. *J. Chem. Educ.* 92: 1338-1343.

Casas, L. 2018. Three-dimensional-printing aids in visualizing the optical properties of crystals. *J. Appl. Cryst.* 51: 1-8.

Chakraborty, P. and Zuckermann, R. N. 2013. Coarse-grained foldable physical model of the polypeptide chain. *P. Natl. Acad. Sci.* 110: 13368-13373.

Chen, T. H., Lee, S., Flood, A. H., et al. 2014. How to print a crystal structure model in 3D. *Cryst. Eng. Comm.* 16: 5488-5493.

Chen, T., Egan, P., Stöckli, F. et al. 2015. Studying the impact of incorporating an additive manufacturing based design exercise in a large, first year technical drawing and CAD course. *Proc. ASME International Design Engineering Technical Conference & Computer and Information in Engineering Conference IDETC/CIE, August 2-5, 2015, Boston MA 1-8*

Coakley, M., Hurt, D., Weber, N. 2014. The NIH 3D print exchange: a public resource for bioscientific and biomedical 3D prints. *3D Printing and Additive Manufacturing* 1: 137-140.

Kat Cooper, A. and Oliver-Hoyo, M. T. 2017. Creating 3D Physical Models to Probe Student Understanding of Macromolecular Structure. *Biochem. Molecular Bio. Educ.* 45: 491-500.

Da Veiga Beltrame, E., Tyrwhitt-Drake, J. Roy, I., et al. 2017. 3D Printing of Biomolecular Modes for Research and Pedagogy. *J. Vis. Exp.* 121: e55427 (8 pages)

Davenport, J., Pique, M., Getzoff, E., et al. 2017. A Self-Assisting Protein Folding Model for Teaching Structural Molecular Biology. *Structure* 25: 671-678.

Dean, N. L., Ewan, C., and McIndoe, J. S. 2016. Applying Hand-Held 3D Printing Technology to the Teaching of VSEPR Theory. *J. Chem. Educ.* 93: 1660-1662.

Downs, R. T., Hall-Wallace, M. 2003. The American Mineralogist crystal structure database. *American Mineralogist* 88: 247-250. http://rruff.geo.arizona.edu/AMS/amcsd.php, assessed May 14, 2018.

Flint, E. B. 2011. Teaching Point-Group Symmetry with Three-Dimensional Models. *J. Chem. Educ.* 88: 907-909.

Ford, S. Minshall, T. 2018. Invited review article: Where and how 3D printing is used in teaching and education. *Additive Manufacturing* 25: 131-150.

Fuentes-Cobas, L. E. 2017a. http://mpod.cimav.edu.mx/, assessed May 14, 2018.

Fuentes-Cobas, L. E., Chateigner, D., Fuentes-Montero, M. E., et al. 2017b. The representation of coupling interactions in the Material Properties Open Database (MPOD). *Advances in Applied Ceramics* 116: 428-433.

Gardner, A. and Olsen, A. 2016. 3D printing of molecular models. *J. Biocommun.* 40: 15-21.

Gartner press release: Gartner Says Consumer 3D Printing Is More Than Five Years Away, Egham, U.K., August 19, 2014, http://www.gartner.com/newsroom/id/2825417. (Assessed April 24, 2018.)

Gatto, A., Bassoli, E., and Denti, L. 2015. Multi-disciplinary approach in engineering education: learning with additive manufacturing and reverse engineering. *Rapid Prototyping Journal* 21: 598-603.

Gillet, A., Sanner, M., Stoffler, D., et al. 2005. Tangible Interfaces for Structural Molecular Biology. *Structure* 13: 483-491.

Glazer, A. M. 2018. Confusion over the description of the quartz structure yet again. *J. Appl. Cryst.* 51: 915-918.

Gražulis, S., Chateigner, D., Downs, R. T., et al. 2009. Crystallography Open Database – an open access collection of crystal structures, *J. Appl. Cryst.* 42: 726-729.

Gražulis, S., Daškevič, A., Merkys, A. et al. 2011. Crystallography Open Database (COD): an open-access collection of crystal structures and platform for world-wide collaboration, *Nucleic Acids Research* 40: D420-D427.

Gražulis, S., Sarjeant, A. A., Moeck, P., et al. 2015. Crystallographic Education in the 21$^{st}$ Century. *J. Appl. Cryst.* 48: 1964-1975.

Gražulis, S., Merkys, A., Vaitkus A., et al. 2018. Crystallography Open Database: history, development, perspectives, *Materials Informatics: Methods, Tools and Applications*, In: Wiley Materials Informatics, Wiley-VCH Verlag, GmbH & Co. KGaA, pp. 1-39.

Griffith, K. M., de Cataldo, R., and Fogarty, K. H. 2016. Do-It-Yourself: 3D Models of Hydrogenic Orbitals through 3D Printing. *J. Chem. Educ.* 93: 1586-1590.

Groom, C. R., Bruno, I. J., Lightfoot, M. P., et al. 2016. The Cambridge Structural Database. *Acta Crystallogr. B 72*: 171-179.

Gross, B. C., Erkal, J. L., Lockwood, S. Y. et al. 2014. Evaluation of 3D Printing and its Potential Impact on Biotechnology and the Chemical Sciences. *Analytical Chemistry* 86: 3240-3253.

Halford, B. 2014. 3-D Models, Without the Kit. *Chemical and Engineering News (C&EN)* 92: 32-33.

Hall, S. R., Allen, F. H., and Brown, I. D. 1991. The crystallographic information file (CIF): a new standard archive file for crystallography. *Acta Cryst. A* 47: 655-685.

Hall, Sidney, R., McMahon, Brian, (eds.). 2005. International Tables for Crystallography, Vol. G: Definition and exchange of crystallographic data. 1$^{st}$ edition. International Union of Crystallography & Wiley.

Hanson, R. M. 2010. Jmol – a paradigm shift in crystallographic visualization. *J. Appl. Cryst.* 43: 1250-1260.

Hanson, R. M. 2016a. JSmol Crystal Symmetry Explorer, https://chemapps.stolaf.edu/jmol/jsmol/jcse/explore.htm.

Hanson, R. M. 2016b. Jmol 3D print, http://chemapps.stolaf.edu/jmol/3dprint/, assessed May 14, 2018.

Hanson, R. M. 2016c. https://sourceforge.net/projects/jmol/?source=navbar, assessed May 14, 2018.

Hart, G. W. 2005. Creating a Mathematical Museum on your desk. *Mathematical Intelligencer* 27: 14-17.

Herman, T., Morris, J., Colton, S., et al. 2006. Tactile Teaching: Exploring Protein Structure/Function using Physical Models. *Biochem. Mol. Biol. Edu.* 34: 247-254.

Higman, C. S. Situ, H, Blacklin, P. et al. 2017. Hands-On Data Analysis: Using 3D Printing to Visualize Reaction Progress Surfaces. *J. Chem. Educ.* 93: 1586-1590.

Horowitz, S. S. and Schultz, P. H. 2014. Printing Space: Using D Printing of Digital Terrain Models in Geosciences Education and Research. *J. Geoscience Educ.* 62: 138-145.

Jittivadhna, K., Ruenwongsa, P., and Panjipan, B. 2010. Hand-held models of ordered DNA and protein structures as 3D supplements to enhance student learning of helical biopolymers. *Biochem. Mol. Biol. Edu.* 38: 359-364.

Jones, O. A. and Spencer, M. J. S. 2018. A Simplified Method for the 3D Printing of Molecular Models for Chemical Education. *J. Chem. Edu.* 95: 88-96.

Kaliankin, D. S., Zaari, R. R., and Varganov, S. A. 2015. 3D Printed Potential and Free Energy Surfaces for Teaching Fundamental Concepts in Physical Chemistry. *J. Chem. Educ.* 92: 2106-2112.

Kaminsky, W. 2000. Wintensor: ein WIN95/98/NT Programm zum Darstellen tensorieller Eigenschaften. *Zeitschrift für Kristallographie Suppl.* 17: 51.

Kaminsky, W. 2005. WinXMorph: a computer program to draw crystal morphology growth sectors and cross sections with export files in VRML V2.0 utf8-virtual reality format *J. Appl. Cryst.* 38: 566-567.

Kaminsky, W. 2007. From CIF to virtual morphology using the WinXMorph program *J. Appl. Cryst.* 40: 382-385.

Kaminsky, W., Snyder, T., Stone-Sundberg, J., et al. 2014. One-click preparation of 3D print files (*.stl *.wrl) from *.cif (crystallographic information framework) data using Cif2VRML. *Powder Diffraction* 29: S42-S47.

Kaminsky, W., Snyder, T., Stone-Sundberg, J., et al. 2015. 3D printing of representation surfaces from tensor data of $KH_2PO_4$ and low-quartz utilizing the WinTensor software. *Zeitschrift für Kristallographie – Crystalline Materials* 230: 651- 656.

Kitson, P., Macdonell, A., Tsuda, S., et al. 2014. Bringing Crystal Structures to Reality by Three-Dimensional Printing. *Crystal Growth & Design* 14: 2720-2724.

Lolur, P. and Dawes, R. 2014. 3D Printing of Molecular Potential Energy Surface Models, *J. Chem. Educ.* 91: 1181-1184.

Macrae, C. F., Bruno, I. J., Chisholm, J. A., et al. 2008. Mercury CSD 2.0 - New features for the visualisation and investigation of crystal structures. *J. Appl. Crystallogr.* 41: 466-470.

Meyer, S. C. 2015. 3D Printing of Protein Models in an Undergraduate Laboratory: Leucine Zippers. *J. Chem. Educ.* 92: 2120-2125.

Moeck, P. 2004. http://nanocrystallography.reserach.pdx.edu, assessed May 14, 2018

Moeck, P., Čertík, O., Seipel, B., et al. 2005. Crystal structure visualizations in three dimensions with database support, *Mater. Res. Soc. Symp. Proc.* 909E: 0909-PP03-05.1-6.

Moeck, P., Čertík, O., Upreti, G., et al. 2006a. Crystal structure visualizations in three dimensions with support from the open access Nano-Crystallography Database, *J. Mater. Educ.* 28: 87-95.

Moeck, P., Seipel, B., Upreti, G., et al. 2006b. Freely Accessible Internet Resources for Nanoscience and Nanotechnology Education and Research at Portland State University's Research Servers, *Mater. Res. Soc. Symp. Proc.* 931: 0931-KK01-04.

Moeck, P. 2006c, www.nanocrystallography.org, assessed May 14, 2018.

Moeck, P., Kaminsky, W., and Snyder, T. J. 2014a. Presentation and answers to a few questions about 3D printing of crystallographic models. *International Union of Crystallography Newsletter* 22: 7-9.

Moeck, P., Stone-Sundberg, J., Snyder, T., et al. 2014b. Crystallography in Interdisciplinary College Education Settings: Educational Offsprings of the Crystallography Open Database and 3D Printed Crystallographic Models. in: *Educating and Mentoring Young Materials Scientists for Career Development*, Liu. L. et al. (eds.). *Mater. Res. Soc. Symp. Proc.* 1716: mrss14-1716-fff03-09 (6 pages).

Moeck, P., Stone-Sundberg, J., Snyder, T. J., et al. 2014c. Enlivening a 300 level general education class on nanoscience and nanotechnology with 3D printed crystallographic models. *J. Mater. Edu.* 36: 77-96.

Moeck, P., Kaminsky, W., Fuentes-Cobas, L., et al. 2016a. 3D printed models of materials tensor representations and the crystal morphology of alpha quartz. *Symmetry: Culture and Science* 27: 319-330.

Moeck, P. 2016b. http://nanocrystallography.reserach.pdx.edu/3D-print-files/, assessed May 14, 2018.

Moeck, P., DeStefano, P., Cheung, I., et al. 2017a. Straightforward routes from CIFs to three-dimensional printed crystallographic models, *Acta Cryst.* A 73: C1132.

Moeck, P. 2017b. http://nanocrystallography.research.pdx.edu/3d-print-files/convert/, assessed May 14, 2018.

Moeck, P. and Snyder T. 2018. Preparing 3D print files for nano-tech/science education from entries of large open-access crystallographic databases at dedicated websites. Proc. 2018 IEEE 13th Nanotechnology and Devices Conference, October 14-17, 2018, Portland, OR, doi: 10.1109/NMDC.2018.8605880.

Nadel, B. 2017. 3D printer capabilities form the future of higher ed. *University Business* May 29, https://www.universitybusiness.com/article/3d-printer-capabilities-form-future-higher-ed, accessed April 22, 2018.

National Institute of Health, 3D print exchange website, http://3dprint.nih.gov, accessed April 21, 2018.

Olson, A. J., Hu, Y. H. E., and Keinan, E. 2007. Chemical mimicry of viral capsid self-assembly. *Proc. Natl. Acad. Sci.* 104: 20731-20736.

Paukstelis, P. J. 2018. MolPrint3D: Enhanced 3D Printing of Ball-and-Stick Molecular Models, *J. Chem. Educ.* 95: 169-172.

Papazafiropulos, N., Fanucci, L., and Leporini, B. 2016. Haptic Models of Arrays through 3D Printing for Computer Science Education. in: *Computers Helping People with Special Needs.* Lecture Notes in Computer Science. 9758: 491-498.

Piunno, P. A. 2017. Teaching the Operation Principles of a Diffraction Grating Using a 3D-Printable Demonstration Kit. *J. Chem. Educ.* 94: 615-620.

Peinador, C., Blanco, V., and Quintela, J. M. 2009. A New Doubly Interlocked [2]Catenane, *J. Am. Chem. Soc.* 131: 920-921.

Penny, M. R., Cao, Z. J., Patel, B. et al., 2017. Three-Dimensional Printing of a Scalable Molecular Model and Orbital Kit for Organic Chemistry Teaching and Learning. *J. Chem. Educ.* 95: 1265-1271.

Pettersen, E. F., Goddard, T. D., Huang, C. C., et al. 2004. UCSF Chimera a visualization system for exploratory research and analysis. *J. Comp. Chem.* 25: 1605-1612.

Ridley, M. 2017. Don't write off the next big thing too soon. *The Times*, November 6, 2017, https://www.thetimes.co.uk/article/dont-write-off-the-next-big-thing-too-soon-rbf2q9sck, assessed May 17, 2018.

Roberts, J., Hagedorn, E., Dillenburg, P., et al. 2005. Physical models enhance molecular 3D literacy in an introductory biochemistry course. *Biochem. Mol. Biol. Edu.* 33: 105-110.

Robertson, M. J. and Jorgensen, W. L. 2015. Illustrating Concepts in Physical Organic Chemistry with 3D Printed Orbitals. *J. Chem. Educ.* 92: 2113-2116.

Rodenbough, P. P., Vanti, W. B., and Chan, S-W. 2015. 3D-Printing Crystallographic Unit Cells for Learning Materials Science and Engineering. *J. Chem. Educ.* 92: 1960-1962.

Rossi, S., Benaglia, M., Brenna, D., et al. 2015. Three Dimensional (3D) Printing: A Straightforward, User-Friendly Protocol to Convert Virtual Chemical Models to Real-Life Objects. *J. Chem. Educ.* 92: 1398-1401.

Scalfani, V. F. and Vaid, T. P. 2014. 3D Printed Molecules and Extended Solid Models for Teaching Symmetry and Point Groups.. *J. Chem. Educ.* 91: 1174-1180.

Scalfani, V. F., Turner, C. H., Rupar, P. A., et al. 2015. 3D Printed Block Copolymer Nanostructures. *J. Chem. Educ.* 92: 1866-1870.

Scalfani, V. F., Williams, A. J., Tkachenko, V., et al. 2016a. Programmatic conversion of crystal structures into 3D printable files using Jmol. *Journal of Cheminformatics* 8: 66 (8 pages).

Scalfani, V. F. 2016b. https://dx.doi.org/10.6084/m9.figshare.c.3302859.v6, assessed May 14, 2018.

Schwarzenbach, Dieter. 1997. *Crystallography*, John Wiley & Sons.

Segerman, H. 2012. 3D Printing for Mathematical Visualization. *Mathematical Intelligencer* 34: 56-62.

Simpson, T. W., Williams, C. B., Hripko, M., et al. 2017. Preparing industry for additive manufacturing and its applications: Summary & recommendations from the National Science Foundation workshop. *Additive Manufacturing* 13: 166-178.

Snyder, T., Andrews, M., Weislogel, M., et al. 2014. 3D Systems' Technology Overview and New Applications in Manufacturing Engineering Science and Education. *3D Printing and Additive Manufacturing* 1: 169-176.

Smiar, K. and Mendez, J. D. 2016. Creating and Using Interactive, 3D-Printed Models to Improve Student Comprehension of the Bohr Model of the Atom, Bond Polarity, and Hybridization. *J. Chem. Educ.* 93: 1591-1594.

Stoddart, J. F. Mechanically Interlocked Molecules (MIMs) – Molecular Shuttles, Switches, and Machines (Nobel Lecture). 2017. *Angew. Chem. Int. Ed.* 56: 11094-11125.

Stone-Sundberg, J., Snyder, T., Kaminsky, W., et al. 2015. 3D printed models of small and large molecules crystal structures and morphologies of crystals as well as their anisotropic physical properties. *Cryst. Res. Technol.* 50: 432-441.

Striplin, D. R., Blauch, D. N., and Carroll, F. A. 2015. Discovering Pressure-Volume-Temperature Phase Relationships with 3D Models. *Chemical Educator* 20: 271-275.

Tavousi, P., Amin, R., and Shahbazmohamadi, S. 2018. Assemble-and-Match: A Novel Hybrid Tool for Enhancing Education and Research in Rational Structure Based Drug Design. *Scientific Reports* 8: 849 (14 pages).

Teplukhin, A. and Babikov, D. 2015. Visualization of Potential Energy Functions Using an Isoenergy Approach and 3D Prototyping. *J. Chem. Educ.* 92: 305- 309.

Van der Meulen, R. 2016. private communication per email.

Van Wieren, K., Tailor, H. N., Scalfani, V. F., et al. 2018. Rapid Access of Multicolor Three-Dimensional Printed Chemistry and Biochemistry Models Using Visualizations and Three-Dimensional Printing Software Programs. *J. Chem. Educ.* 94: 964-969.

Violante, L. E. H., Nunez, D. A., Ryan, S. M., et al. 2014. 3D Printing in the Chemistry Curriculum: Inspiring millennial students to be creative innovators. *Chapter 9 in: Addressing the Millennial Student in Undergraduate Chemistry* (editors: Potts G. E. and Dockery C. R.), ACS Symposium Series 1180: 125-146.

Wedler, H. B., Cohen, S. R., Davis, R. L., et al. 2012. Applied Computational Chemistry for the Blind and Visually Impaired. *J. Chem. Educ.* 89: 1400-1404.

Wikipedia 2018. Hype cycle, https://en.wikipedia.org/wiki/Hype_cycle, assessed April 19, 2018.

Wood, P. A., Sarjeant, A. A., Bruno, I., et al. 2017. The next dimension of structural science communication: simple 3D printing directly from a crystal structure, *CrystEngComm* 19: 690-698.